# A Probabilistic Approach to Analysis of Reliability in *n*-D Meshes with Interconnect Router Failures

Farshad Safaei and Majed ValadBeigi

Faculty of ECE, Shahid Beheshti University G.C., Evin 1983963113, Tehran, IRAN

f_safaei@sbu.ac.ir, m.valadbeigi@mail.sbu.ac.ir

*Abstract*

*The routing algorithms for parallel computers, on-chip networks, multi-core processors, and multiprocessors system-on-chip (MP-SoCs) exhibit router failures must be able to handle interconnect router failures that render a symmetrical mesh non-symmetrically. When developing a routing methodology, the time complexity of calculation should be minimal, and thus complicated routing strategies to introduce profitable paths may not be appropriate. Several reports have been released in the literature on using the concept of fault rings to provide detour paths to messages blocked by faults and to route messages around the fault regions. In order to analyze the performance of such algorithms, it is required to investigate the characteristics of fault rings. In this paper, we introduce a novel performance index of network reliability presenting the probability of message facing fault rings, and evaluating the performance-related reliability of adaptive routing schemes in n-D mesh-based interconnection networks with a variety of common cause fault patterns. Sufficient simulation results of Monte-Carlo method are conducted to demonstrate the correctness of the proposed analytical model.*

*Keywords*

*Fault-Tolerance, Fault Rings, Interconnection Networks, n-D Meshes, Adaptive Routing, Network Reliability, Performance Analysis*

## 1. Introduction

In the recent years, high performance computing systems have opened a trend in modeling daily behavior and life style of modern societies by means of applications and services, such as molecular dynamics simulations, DNA sequencing, weather forecasting, geological activity studies, and even a simple Google search [1].

In massively parallel systems, a collection of components work together to solve problems in large-scale applications. These components communicate data, and coordinate their effort by sending and receiving messages through the underlying network of communications. However, as the number of components in a parallel system increases, so does the rate of failure. Therefore, the issue of system reliability must acquire an increased importance.

Fault-tolerance is the ability of networks to operate in the presence of component failures [2]. It is a key issue in the design of interconnection networks. When components of networks fail, higher adaptability of routing algorithm would enable message to be routed around fault regions.

A wide range of researches have been carried out on fault-tolerant routing in interconnection networks [3-10]. Typically, additional routing restrictions and/or network resources are required to ensure deadlock freedom of such algorithms in the vicinity of faults [2]. Furthermore, faults may be coalesced into some *fault areas* that engage rectangle-shaped (also called fault- block) regions such that the boundary of the rectangle has only fault-free nodes and links and the interior of the rectangle contains all the faulty links and nodes which

DOI : 10.5121/ijcnc.2011.3407                                                              87

correspond to that particular fault set. This fault model is superior to convex fault regions proposed by Chien and Kim [11], because it deals more effectively with faults along the network boundary. For each fault region in a network with faults, it is feasible to connect the fault-free components around the region to form a ring known as the *fault ring* (f-ring for short) for that fault.

Many researchers [3-7] have utilized the concept of *f-rings* in order to make the existing routing algorithms tolerate multiple fault regions without disabling any healthy nodes. Messages are routed under an adaptive routing algorithm until they face an f-ring. Then, depending on the relative position of the destination, messages are routed clockwise or counterclockwise around the f-ring. The f-rings provide alternate paths to messages blocked by faults which are used to route messages around block-fault regions. An f-ring can be constructed for each rectangular fault region by using the methods as described in [3]. Consequently, in order to evaluate the end-to-end cost and complexity of such routing schemes accurately; we ought to be able to investigate the characteristics of the fault rings.

Several studies [3-11] have addressed fault-tolerance in a diverse range of systems and applications. However, most of the performance studies have been performed through numerical simulation model which, were solely based on an analysis of the discrete time behavior of the system. The analytical modeling approach is a better choice when a lot of tests are required. While, numerical simulation approach is time consuming to create, although it can be used to generate quick results. If both simulation and analytical modeling are feasible, then the optimum technique depends on the type of investigations performed.

The *mesh-connected topology* [2] is one of the most thoroughly investigated topologies of networks for on-chip networks, multi-core processors, and parallel systems. This topology has been commonly discussed due to structural regularity for easy construction, and high potential of legibility of various algorithms. As the number of nodes, as well as dimension in a mesh-connected system increases, the chance of failure also increases. The complex nature of networks also makes them vulnerable to disturbances. Therefore, the ability to tolerate failures is becoming increasingly important for adaptive routing schemes.

However, graphs generally provide a powerful tool to model objects and relationships among objects. For example they can be used to model problems in areas such as transportation, scheduling, networks, robotics, VLSI, compilers, mathematical biology and software engineering. Many optimization problems from such areas and other diverse areas can be phrased in graph-theoretic terms, leading to algorithmic questions about graphs.

In this paper, the focus is on a novel mathematical model for analyzing the reliability of a mesh-based interconnection network which involves probability theory and combinatorics, and moreover, it is used as a simulator for results verification.

In Section 2, all required definitions and backgrounds of our analysis are given. In Section 3, we derive in detail the probability of message facing the fault rings in *n*-D mesh-based networks. In Section 4, mathematical expressions of this probability are confirmed by a Monte-Carlo simulation approach. Finally, in Section 5, our conclusions and anticipated future works are presented.

## 2. Preliminaries

This section briefly describes the structure of an *n*-D mesh network and then explores the fault model used in this study. The definitions brought in this section stay in firm with the standard notation and definitions in adaptive routing algorithms and fault-tolerant networks.

### 2.1. The Mesh Structure

**Definition 1**: Each mesh network of $n$ dimensions $R_1 \times R_2 \times ... \times R_n$ which is denoted by $M_{R_1 \times ... \times R_n}$ is made of nodes $v(M_{R_1 \times ... \times R_n}) = \{(x^1,...,x^n) : 1 \leq x^n \leq R_i, 1 \leq i \leq n\}$ where each node $(x^{i_1},...,x^{i_n})$ is connected to its neighbors $(x^{i_1} \pm 1, x^{i_2},...,x^{i_n})$, $(x^{i_1}, x^{i_2} \pm 1, x^{i_3},...,x^{i_n})$,..., $(x^{i_1},...,x^{i_{n-1}}, x^{i_n} \pm 1)$ if the neighbors exist. Generally, there are $N = R_1 \times ... \times R_n$ nodes and

$$E = nR_1...R_n - \sum_{1 \leq i_1,...i_{n-1} \leq n} R_{i_1}...R_{i_{n-1}} + \sum_{1 \leq i_1,...i_{n-2} \leq n} R_{i_1}...R_{i_{n-2}} - ... + (-1)^{n-1} \sum_{i=1}^{n} R_1$$
$$= \sum_{j=1}^{n} (-1)^{n-(j-1)} \sum_{1 \leq i_1,...i_j \leq n} R_{i_1}...R_{i_j} \qquad (1)$$

communication channels (links) in a $M_{R_1 \times ... \times R_n}$.

## 2.2. Fault Model

Asynchronous, distributed parallel systems, which contain no central controller, are becoming a viable alternative in the design of powerful computing systems. In addition to their characteristics for performance scalability, the physical replication of multiple entities makes them very applicable to the implementation of some kind of fault-tolerance, in which the whole system is allowed to continue operating, albeit in a degraded mode, in the presence of some faulty components.

In designing a fault-tolerant routing scheme, the most important issues are to select appropriate fault models, structure of fault regions, and processes to component failures.
The nodes of network are generally assumed to possess the ability for self-test, as well as the ability to test neighboring nodes. Two basic fault models could be tested, including individual link failures and node failures. On a node failure, all physical links incident on the failed router are also marked faulty at adjacent routers. Link or node failures do not only reduce the computational power of underlying network, but also deform its structure, which may accordingly lead to a *disconnected* network.

**Definition 2:** *A topology is said to be connected if there exists a path between all source-destination pairs* $(s,d)$ *for all* $s \neq d$. *Otherwise, it is disconnected.*

Individual link and node failures are the smallest unit of failed components. Adjacent faulty entities are coalesced into fault regions, which may lead to different patterns of failed entities. Fault regions extended by faulty components, may form *convex* (also known as block faults) or *concave*-shaped patterns [12, 13].

Since concave regions are very difficult to process for fault-tolerant routings, constraints must be placed on the structure of such regions. The most common constraint employed is that these regions be convex and further rectangle-shaped fault regions.

## 3. The Proposed Analytical Model

This section starts with the description of the assumptions used in construction of the analytical model. The derivation and implementation procedure of the mathematical expressions are then presented. After that, the proposed analytical model is validated through simulation experiments.

### 3.1. Assumptions

The topological structure of an interconnect network can be modeled by a graph. This fact has been universally accepted and used by computer scientists and engineers. Moreover, practically it has been demonstrated that graph theory is a very powerful mathematical tool for designing and analyzing topological structure of interconnect networks. In this section, we will briefly recall some basic concepts and notations of graph theory used in this paper as well as the corresponding backgrounds of networks.

1. Messages are uniformly destined to other network nodes.
2. Messages are routed adaptively using the minimal routing, which always routes the message to the destination through a Manhatan-distance-path [9] in the fault-free segments of the network.
3. Both source and destination routers are fault-free.
4. The fault model is static, i.e., no new faults occur during a routing process; faults do not disconnect the network.
5. Nodes are more complex than links and thus have higher failure rates [2, 12, 13]. Thus, we assume only node failures. Moreover, the faulty components are unusable, that is, data will not be transmitted over the faulty entities.

Most of the mathematical models investigated in fault- tolerant communication networks are based on the general assumptions made above.

## 3.2. Mathematical Analysis

In this section, we calculate the probability of message facing the fault rings in $n$-D meshes. Consider an $n$-D $R_1 \times ... \times R_n$ mesh network with some faulty nodes that are put by one another in an $n$-D $l_1 \times l_2 \times ... \times l_n$ ($1 \leq l_i \leq R_i - 1, 1 \leq i \leq n$) rectangular. It is clear that faulty nodes will not cause disconnection in the given $n$-D mesh network. Such a network is called a connected $n$-D $R_1 \times ... \times R_n$ mesh with $n$-D $l_1 \times l_2 \times ... \times l_n$ rectangular fault pattern.

**Definition 3:** *Consider a connected $n$-D $R_1 \times ... \times R_n$ mesh with $n$-D $l_1 \times l_2 \times ... \times l_n$ rectangular fault pattern. A set of nodes that have a distance of a link from border points of this fault pattern is called n-D fault ring or $f^n$-ring for short.*

**Definition 4:** *If the given n-D fault pattern meets the borders of n-D mesh, then the n-D fault pattern in Definition 3 changes to n-D fault chain or $f^n$-chain for short.*

For instance, Figure 1 illustrates a part of a 3-D faulty mesh, in which nodes and links outside the cube are omitted, but $f^3$-ring and $f^3$-chain are drawn in. The dimensions are labeled $X^1$ (first), $X^2$ (second), and $X^3$ (third).

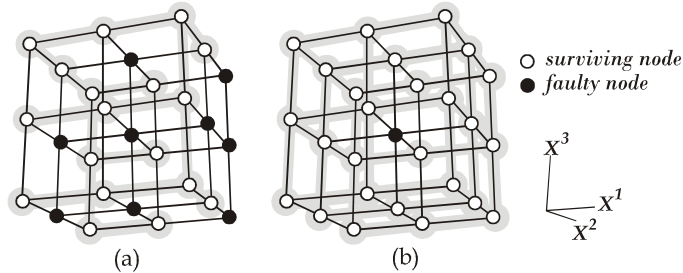

Figure 1. 3-D faulty mesh network; (a) $f^3$-chain, (b) $f^3$-ring

**Definition 5:** Minimal path from point $a = (x_a^1,...,x_a^n)$ to point $b = (x_b^1,...,x_b^n)$: *We denote $S$ as the set of all paths that start from $a$ and terminate at $b$. Let $s_0$ be the length of the shortest sequence of $S$. Therefore, one minimal path from $a$ to $b$ is an element of the set $S$ having the length of $s_0$.*

**Definition 6:** *The distance between the i-th components of points $a = (x_a^1,...,x_a^n)$ and $b = (x_b^1,...,x_b^n)$ in an $n$-D $R_1 \times ... \times R_n$ mesh network equals $|x_b^i - x_a^i|$, which is depicted by symbol $\Delta_i(a,b)$.*

**Definition 7:** The symbol $LT^n(a,b)$ denotes the number of paths (minimal) between two non-faulty points $a,b \in v(M_{R_1 \times ... \times R_n})$ and its value is given by

$$\begin{pmatrix} \sum_{i=1}^{n} \Delta_i(a,b) \\ \Delta_1(a,b),...,\Delta_n(a,b) \end{pmatrix} = \frac{(\Delta_1(a,b)+...+\Delta_n(a,b))!}{\Delta_1(a,b)!\Delta_2(a,b)!...\Delta_n(a,b)!} \quad (2)$$

**Remark:** To reach the point $b$ from the point $a$, we should pass $\Delta_1(a,b)+...+\Delta_n(a,b)$ links. In each movement we can move through the direction of any coordinate axis but on direction of axis $x^i$, we can only choose $\Delta_i(a,b)$ channels. So, the number of minimal paths from $a$ to $b$ is given by

$$\frac{(\Delta_1(a,b)+...+\Delta_n(a,b))!}{\Delta_1(a,b)!...\Delta_n(a,b)!} \quad (3)$$

Our intention is to calculate the probability of an arbitrary path between two non-faulty nodes in an *n*-D mesh not facing $f^n$-*ring*, which is denoted by $P_{miss}$. Hence, our sample space is the set of all minimal paths between two arbitrary non-faulty nodes in the network. Some paths in this set face $f^n$-*ring* while the others do not touch any nodes of the $f^n$-*ring*, so

$$P_{miss} \quad \frac{\text{\# all minimal paths between two arbitrary non-faulty nodes not facing the } f^n\text{-ring}}{\text{\# all minimal paths between two arbitrary non-faulty nodes in the netwok}} \quad (4)$$

and

$$P_{hit} \quad \frac{\text{\#all minimal paths between two arbitrary non-faulty nodes facing the } f^n\text{-ring}}{\text{\#all minimal paths between two arbitrary non-faulty nodes in the netwok}} \quad (5)$$

such that, $P_{hit} = 1 - P_{miss}$.

Note that $P_{miss}$ calculates the probability of an arbitrary minimal path in the network not facing the $f^n$-*ring*. It is apparent that $P_{miss}$ is not the probability of an arbitrary path between two certain nodes of $a,b$ not facing the set of nodes residing on $f^n$-*ring*.

As a matter of fact $P_{miss}$ determines the probability of an arbitrary path between two arbitrary nodes in the network not facing the $f^n$-*ring*, and the aim of this paper is to calculate the probability of a path in the network (without considering the initial and end points of the path) not facing $f^n$-*ring*. Indeed, this probability is equal for all paths existing in the network.

In the *n*-D mesh, the location of the faulty points is important. So, in order to characterize the exact location of faulty points we should determine one of the points of the *n*-D rectangular fault. For the adaptation, we choose a point where for each $1 \le i \le n$, its *i*-th component has the least value among the *i*-th components of the set of the faulty points of the *n*-D rectangular and it is depicted by $\alpha$. Therefore, in order to determine the exact location of the fault pattern of the *n*-D rectangular, which is important in calculation of $P_{hit}, P_{miss}$, apart from knowing dimensions of the n-D rectangular fault we should know the characteristic point of it, i.e. $\alpha$. Thus, the fault pattern of *n*-D rectangular $l_1 \times l_2 \times ... \times l_n$ is as following set of nodes

$$\{(x_a^1 + m_1,...,x_a^n + m_n) / \alpha = (x_a^1,...,x_a^n), 0 \le m_i \le l_i, 1 \le i \le n\} \quad (6)$$

The above set is demonstrated by symbol $F(l_1,...,l_n,\alpha)$, and also the set of $f^n$-*ring* point surrounding this fault pattern is depicted by $R(l_1,...,l_n,\alpha)$. The set of $F(l_1,...,l_n,\alpha) \cup R(l_1,...,l_n,\alpha)$ is illustrated by $FR(l_1,...,l_n,\alpha)$.

**Theorem 1:** *The number of all existing paths among non-faulty nodes in n-D $R_1 \times ... \times R_n$ mesh with n-D $l_1 \times l_2 \times ... \times l_n$ rectangular fault pattern and characteristic point $\alpha$ is given by*

$$\sum_{a,b \in v(M_{R_1 \times ... \times R_n}) \setminus F(l_1,...,l_n,\alpha)} \binom{\sum_{i=1}^{n} \Delta_i(a,b)}{\Delta_1(a,b),...,\Delta_n(a,b)} \tag{7}$$

***Proof*:** Consider two none-faulty points $a,b$ in the network; the number of paths between $a,b$ is denoted by

$$\binom{\sum_{i=1}^{n} \Delta_i(a,b)}{\Delta_1(a,b),...,\Delta_n(a,b)} = LT^n(a,b) \tag{8}$$

So, the number of all existing paths among none-faulty points in the n-D $R_1 \times ... \times R_n$ mesh can be calculated as the aggregation of total number of paths between any two non-faulty points in the network. ∎

To continue the work, it is required to determine the direction of each path in the mesh relative to the coordinate axis. Consider two points $a = (x_a^1,...,x_a^n), b = (x_b^1,...,x_b^n)$ in the *n*-D $R_1 \times ... \times R_n$ mesh network. Assume that we want to move from $a$ to $b$. Table 1 gives the information for the direction of the message utilizing the $X^1,...,X^n$ dimensions in the above mentioned network. The upper indices "+", "−" reveal the orientation in the negative or positive direction of the coordinate axis, respectively.

**Definition 8:** *If the set of faulty points in an n-D $R_1 \times ... \times R_n$ mesh forms two n-D $l_1 \times l_2 \times ... \times l_n$ and, $k_1 \times k_2 \times ... \times k_n$ rectangular-shaped regions with an $f^n$-ring around them, then we will denote the set of fault points by $F_{overlap}$ and the set of fault points and its associated $f^n$-ring around it by $FR_{overlap}$.*

TABLE 1. Description of the direction to be used for routing messages in *n*-D mesh network.

| Condition | Message Direction |
|---|---|
| $x_b^i - x_a^i \geq 0$ | In $X^i$ dimension the message is routed from $a = (x_a^1,...,x_a^n)$ to $b = (x_b^1,...,x_b^n)$ in $X^{i+}$ direction |
| $x_b^i - x_a^i < 0$ | In $X^i$ dimension the message is routed from $a = (x_a^1,...,x_a^n)$ to $b = (x_b^1,...,x_b^n)$ in $X^{i-}$ direction |

It is apparent that in this case we should have $\alpha_1$ and $\alpha_2$ apart from the dimensions of the faulty rectangular-shaped regions. If the set of considered faulty points does not have a specific shape (i.e., $x-shape$), we denote that set of faulty points $F(x)$ and $FR(x)$ is called the set of fault points and its associated $f^n$-ring around it, respectively.

**Definition 9:** *Restriction function of $M_{R_1 \times ... \times R_n}$, the function $\Re : P(M_{R_1 \times ... \times R_n}) \to P(M_{R_1 \times ... \times R_n})$, in which $P(M_{R_1 \times ... \times R_n})$ is the set of all meshes of $M_{R_1 \times ... \times R_n}$ and its criterion is expressed as follows*

$$\Re(M^n(a,b)) = \begin{cases} M^n(a,b) \cap FR(l_1,...,l_n,\alpha), \text{if set of faulty points is in the form of} \\ \text{an } l_1 \times l_2 \times ... \times l_n \text{ rectangle } M^n(a,b) \cap FR_{overlap}, \\ \text{if set of faulty points is in the form of two overlapping rectangles} \\ M^n(a,b) \cap FR(x), \\ \text{if set of faulty points is in the form of a x-shape} \end{cases} \tag{9}$$

**Definition 10:** *Let $a$ and $b$ be two non-faulty points of $v$ ($M_{R_1 \times ... \times R_n}$) and for any two arbitrary points $C_l = (x_{c_l}^1, ..., x_{c_l}^n)$, $C_k = (x_{c_k}^1, ..., x_{c_k}^n)$ of $\Re(M^n(a,b)) \cup \{a,b\}$, the number of possible paths from $C_l$ to $C_k$ is indicated by $LM_{a,b}(C_l, C_k)$; so that the direction of each path in dimension $X^i (1 \le i \le n)$ is collinear with the direction of a path from $a$ to $b$ in dimension $X^i (1 \le i \le n)$, and its value is given by*

$$\left( \begin{array}{c} \sum_{i=1}^{n} \Delta_i^{a,b}(C_l, C_k) \\ \Delta_1^{a,b}(C_l, C_k), ..., \Delta_n^{a,b}(C_l, C_k) \end{array} \right) \tag{10}$$

In which, $\Delta_i^{a,b}(C_l, C_k), 1 \le i \le n$, is a two-criterion function; indicating the number of orientations along path from $C_l$ to $C_k$ in dimension $X^i$, which are collinear with the orientations along a path from $a$ to $b$ in the same dimension. But if the orientations along a path from $C_l$ to $C_k$ in dimension $X^i$, which are not collinear with the orientations along a path from $a$ to $b$, then it gives the negative of the previous value, i.e. its criterion is expressed as

$$\Delta_i^{a,b}(C_l, C_k) = \begin{cases} \Delta_i(C_l, C_k) & x_b^i - x_a^i \ge 0 \text{ and } x_{c_k}^i - x_{c_l}^i \ge 0 \\ & \text{or } (x_b^i - x_a^i < 0 \text{ and } x_{c_k}^i - x_{c_l}^i < 0 \\ -\Delta_i(C_l, C_k) & \text{otherwise} \end{cases} \tag{11}$$

**Definition 11 (Permutation)** [14]: *Given any set $A$, any one-to-one mapping from $A$ onto $A$ is called a non-singular transformation or permutation.*

The symbol $S_k$ will be used to denote the set of all possible permutations of the set $\{0,...,k-1\}$.

**Definition 12 (Identity mapping)** [14]: *Let $A$ be a non-void set and define $i: A \to A$ by $i(x) = x \ \forall x \in A$. In terms of ordered pairs $i = \{(x,x): x \in A\}$, $i$ is called the identity mapping. If $i$ maps at least one element of $A$ into another element excluding itself, then $i$ is said to be a non-identical mapping.*

**Definition 13 (Cycle)** [14]: *Let $A$ be a finite set. A permutation $\sigma$ of $A$ is said to be a cyclic permutation or a cycle if there exist elements $x_1, x_2, ..., x_n$ in $A$ such that $\sigma(x_1) = x_2, \sigma(x_2) = x_3, ..., \sigma(x_n) = x_1$ and for any $x \in A$ that different from $\{x_1, x_2, ..., x_n\}$, $\sigma = (x_1, x_2, ..., x_n)$.*

**Remark:** Every permutation $\sigma$ of a finite set is a product of disjoined cycles.

**Definition 14: (Derangement)** *a permutation $\sigma$ of $S_{m+1}$ is said to be a derangement if it moves all the elements of $\{x_1, x_2, ..., x_m\}$. We denote the set of all parishes in $S_{m+1}$ by symbol $D_{m+1}$ or $D_{\{0,1,...,m\}}$.*

**Definition 15 (Transposition)** [14]: *A cycle of length 2 is a transposition.*

According to definition of determinant, we have

$$\det_{0 \le k, l \le m} d'_{kl} = \sum_{\sigma \in S_{m+1}} (-1)^{\text{sgn}\,\sigma} d'_{0\sigma(0)} d'_{1\sigma(1)} ... d'_{m\sigma(m)} \tag{12}$$

where $S_{m+1}$ and $\text{sgn}\,\sigma$ are the set of all permutations on the set $\{0,1,...,m\}$ and the number of transpositions of permutation $\sigma$, respectively.

**Theorem 2:** *Given that $a,b$ are two non-faulty points of n-D $M_{R_1 \times ... \times R_n}$ mesh network and $\Re(M^n(a,b)) = \{C_1, ..., C_m\}$; the number of path from $a$ to $b$, that do not traverse the points $C_1, ..., C_m$, can be calculated as*

$$\det_{0 \leq k,l \leq m} d_{kl}(a,b) \tag{13}$$

where

$$d_{0l}(a,b) = LM_{a,b}(C_l, C_{m+1}) \quad l = 0,1,...,m$$
$$d_{kl}(a,b) = LM_{a,b}(C_l, C_k)$$
$$k = 1,2,...,m, l = 0,1,...,m, C_{m+1} = b, C_0 = a \tag{14}$$

***Proof*:** In order to make the presentation of our notation easier to understand, let us use symbols $d_{kl}, LM(C_l, C_k)$ in place of $d_{kl}(a,b), LM_{a,b}(C_l, C_k)$, respectively. Let us take $P_{m+1}$ to be the number of all the paths from $a$ to $b$, which do not traverse the points $C_1,...,C_m$. Further, let $S$ be the set of all possible paths from point $a$ to $b$. For any $1 \leq k \leq m, w \in S$ satisfies the condition $t_k$ if the path $w$ passes point $E_k$. Now, let $T_l$ be the number of those elements of $S$, which fulfill $l$ conditions out of all conditions mentioned above. It is clear that $T_0 = \|S\|$, where the symbol $\| \ \|$ indicates the cardinality of set $S$. According to the inclusion–exclusion principle [15], we have

$$P_{m+1} = \sum_{l=0}^{m} (-1)^l T_l \tag{15}$$

By considering those elements of $S$ which satisfy conditions $t_{k_1}, t_{k_2},...,t_{k_l}$, the number of which is equal to the $N(t_{k_1}, t_{k_2},...,t_{k_l})$. Therefore, we get

$$P_{m+1} = LM(C_0, C_{m+1}) - \sum_{i=1}^{m} N(t_l) + \sum_{l<k \neq l \leq m} N(t_k, t_l) - \cdots + (-1)^m N(t_1,...,t_m) \tag{16}$$

The quantity $N(t_{k_1},...,t_{k_l})$ is equal to the number of paths from $S$, which satisfies the conditions $t_{k_1},...,t_{k_l}$; in other words, it equals to paths from $a$ to $b$ such that they traverse all the nodes $C_{k_1},...,C_{k_l}$, and this quantity is expressed as follows;

$$N(t_{k_1},...,t_{k_l}) = \sum_{\sigma \in D_{\{0,l_1,...,l_k\}}} d_{0\sigma(0)} d_{l_1\sigma(l_1)} ... d_{l_k\sigma(l_k)} \tag{17}$$

However, $\sigma \in D_{\{0,l_1,...,l_k\}}$ is equal to a permutation such that $\sigma \in S_{m+1}$, in a way that it is constant on the elements of $\{0,1,...,m\} \setminus \{0,l_1,...,l_k\}$ and $\delta|_{\{0,l_1,...,l_k\}} = \sigma$. Then, we get

$$N(t_{k_1},...,t_{k_l}) = \sum_{\delta} d_{0\delta(0)} ... d_{m\delta(m)} \tag{18}$$

and also we have $\text{sgn}\,\sigma = \text{sgn}\,\delta = k$.

Besides, each permutation such as $\delta \in S_{m+1}$ (that moves only elements of $\{0,l_1,...,l_k\} \subset \{0,...,m\}$) is presented as an extension form of a permutation such that $\delta \in D_{\{0,l_1,...,l_k\}}$, in a way that $\delta|_{\{l_1,...,l_k\}} = \sigma$ and $\delta|_{\{0,1,...,m\} \setminus \{l_1,...,l_k\}}$: const

which completes the proof. ∎

Now, in order to calculate $P_{miss}$, it is necessary to enumerate the number of all paths that do not face the $f^n$-*ring* and then divide them by the number of all existing paths in the network.

**Theorem 3**: *Let $M_{R_1 \times ... \times R_n}$ be a connected n-D $R_1 \times ... \times R_n$ mesh network with the n-D $l_1 \times ... \times l_n$ rectangular fault region and the characteristic point $\alpha$, then the number of paths not crossing the $f^n$-ring is given by*

$$\sum_{a,b \in v(M_{R_1 \times \ldots \times R_n}) \setminus FR(l_1,\ldots,l_n,\alpha)} \det_{0 \leq k,l \leq C_{a,b}} d_{kl}(a,b) \quad (19)$$

in which

$$C_{a,b} = \|\Re(M^n(a,b))\|$$

**Proof:** Considering the fact that each path confronting $F(l_1,\ldots,l_n,\alpha)$ will also be incident with the $f^n$-*ring*, the number of paths not traversing the points located on $f^n$-*ring* is equal to the number of paths in $M_{R_1 \times \ldots \times R_n}$ not crossing the $FR(l_1,\ldots,l_n,\alpha)$ points. Consider two arbitrary points $a$ and $b$ from the set $v(M_{R_1 \times \ldots \times R_n}) \setminus FR(l_1,\ldots,l_n,\alpha)$. According to Theorem 2, the number of minimal paths (from $a$ to $b$) not traversing the points $\Re(M^n(a,b))$, equals $\det_{0 \leq k,l \leq C_{a,b}} d_{kl}(a,b)$. Thus, the number of all existing paths in $M_{R_1 \times \ldots \times R_n}$ not facing the $f^n$-*ring* will be equal to the total number of paths between any two non-faulty points in $M_{R_1 \times \ldots \times R_n}$ not traversing the points of $FR(l_1,\ldots,l_n,\alpha)$.

That is,

$$\sum_{a,b \in v(M_{R_1 \times \ldots \times R_n}) \setminus FR(l_1,\ldots,l_n,\alpha)} \det_{0 \leq k,l \leq C_{a,b}} d_{kl}(a,b) \qquad \blacksquare$$

It follows from the proceeding theorem, the probability that a path in $M_{R_1 \times \ldots \times R_n}$ with the *n*-D $l_1 \times \ldots \times l_n$ rectangular fault pattern and characteristic point $\alpha$ is not blocked by its associated $f^n$-*ring* can be determined as

$$P_{miss} = \frac{\sum_{a,b \in v(M_{R_1 \times \ldots \times R_n}) \setminus FR(l_1,\ldots,l_n,\alpha)} \det_{0 \leq k,l \leq C_{a,b}} d_{kl}(a,b)}{\sum_{a,b \in v(M_{R_1 \times \ldots \times R_n}) \setminus F(l_1,\ldots,l_n,\alpha)} LT^n(a,b)} \quad (20)$$

Therefore, we can get $P_{hit}$ from the expression $P_{hit} = 1 - P_{miss}$ to calculate $P_{miss}$ with overlapping fault rings, so we have

$$P_{miss} = \frac{\sum_{a,b \in v(M_{R_1 \times \ldots \times R_n}) \setminus FR_{overlap}} \det_{0 \leq k,l \leq C_{a,b}} d_{kl}(a,b)}{\sum_{a,b \in v(M_{R_1 \times \ldots \times R_n}) \setminus F_{overlap}} LT^n(a,b)} \quad (21)$$

and also in a condition that fault pattern is in the *x*-shaped fault region, we get

$$P_{miss} = \frac{\sum_{a,b \in v(M_{R_1 \times \ldots \times R_n}) \setminus FR(x)} \det_{0 \leq k,l \leq C_{a,b}} d_{kl}(a,b)}{\sum_{a,b \in v(M_{R_1 \times \ldots \times R_n}) \setminus F(x)} LT^n(a,b)} \quad (22)$$

## 4. Results of Monte-Carlo Simulation

In the previous section, we derived mathematical expressions to calculate the probability of message facing the f-rings (f-chains) with and without overlapping. In this section, a simulation is also constructed and applied under the same assumptions and network conditions in order to validate the results given by the proposed analytical model. The simulation has been developed in the method of Monte-Carlo, that is, we simply take many statistics for every run, independently, without waiting for the number of simulation runs is adjusted to ensure a steady-state operation condition for the network.

International Journal of Computer Networks & Communications (IJCNC) Vol.3, No.4, July 2011Table 2. Analytical model with computer simulation results of the probability of message facing fault rings (chains) in a mesh topology with different characteristics of fault rings (chains), and various dimensions of the network

| Network dimension | Fault region characteristic | Origin of the faulty points set | $n$-D mesh fault | $P_{hit}$ |
|---|---|---|---|---|
| 3×2×2 | $f^3$-chain | (1,1,0) | 1×1×1 | 1 |
| 7×8×11 | $f^3$-ring | (2,2,2) | 2×1×3 | 0.214 |
| 5×13×9 | $f^3$-ring | (2,3,1) | 1×7×2 | 0.304 |
| 3×5×7 | $f^3$-chain | (0,0,1) | 2×2×2 | 0.817 |
| 6×11×17 | $f^3$-ring | (2,4,6) | 4×6×10 | 0.884 |
| 3×7×8×9 | $f^4$-ring | (1,1,1,1) | 1×5×6×8 | 0.878 |
| 2×3×4×2 | $f^4$-chain | (0,0,1,0) | 1×1×2×1 | 0.976 |
| 9×5×3×9 | $f^4$-ring | (2,3,1,3) | 1×1×1×4 | 0.095 |
| 3×3×3×3×3 | $f^5$-chain | (1,1,1,1,1) | 1×1×1×2×1 | 1 |
| 5×4×3×5×6 | $f^5$-ring | (1,1,1,1,1) | 1×1×1×2×1 | 0.036 |
| 5×4×3×5×6 | $f^5$-ring | (2,2,2,2,2) | 2×2×1×1×3 | 0.104 |

The simulator generates faults in the network so that the resulting fault regions are convex. It also checks that all nodes in the network are still connected using adaptive routing algorithm. The goal of the simulation is to calculate the values of the probability of message facing the f-rings (f-chains) for different number of faulty nodes in an $n$-D mesh topology. For every run, the simulator creates the corresponding rectangle-shaped fault pattern, and keeps statistics of the following data:

- The number of minimal paths crossing the network.
- The number of minimal paths confronting the f-rings (f-chains).
- For each source-destination pair, the probability of message facing the f-rings (f-chains) is calculated.

Table 2 shows the results for various sizes of $n$-D meshes with different characteristics of the fault regions. Here, both the analytical and simulation results are shown.

## 5. Conclusions

Due to the exponential growth of circuit integration, including on-chip networks and multi-core processors, in addition to the stochastic nature of communication between different components on a chip, some interconnection networks may show a significant higher failure rates and a Mean Time Between Failures (MTBF) smaller than the execution time of some applications. This means that failures may occur during their executions. Hence, the performance of such systems is heavily dependent on the routing schemes with the ability to operate with router failures. In order to analyze the performance of such routing schemes, one must investigate the characteristics of fault rings. In this paper, we have analyzed the reliability of $n$-D mesh-based interconnection networks as a probability of message facing the fault rings. Working properly with any hardware at hand makes the method quite economical with considerable dependability. We have also conducted extensive simulation experiments in Monte-Calro-method with a various number of faults, the results of which are used to confirm the accuracy of the proposed analytical model. In our future work, we will extend our approach to investigate dynamic networks, in which every entity may become faulty during the routing process. As a result, the fault regions can change their shapes dynamically, and the corresponding fault rings will be adjusted frequently.

**Authors**

**Farshad Safaei** received the B.Sc., M.Sc., and Ph.D. degrees in Computer Engineering from Iran University of Science and Technology (IUST) in 1994, 1997 and 2007, respectively. He is currently an assistant professor in the Department of Electrical and Computer Engineering, Shahid Beheshti University, Tehran, Iran. His research interests are performance modelling/evaluation, Interconnection networks, computer networks, and high performance computer architecture.

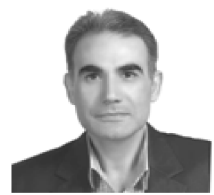

**Majed ValadBeigi** received the B.Sc. degree in Computer Hardware Engineering from Shahid Rajaee University, Tehran, Iran, in 2008. He is currently M.Sc. student in Computer Architecture Engineering in Department of Electrical and Computer Engineering, Shahid Beheshti University, Tehran, Iran. He is a member of Scorpius simulation team which has participated in many international RoboCup Competitions. He is also an IEEE student member. His main Researches focus on network-on-chips, interconnection networks, and performance evaluation.

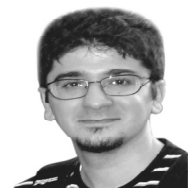